\documentclass[aps,prl,reprint,superscriptaddress,preprintnumbers,showpacs,showkeys]{revtex4-1}

\usepackage{amssymb}
\usepackage{amsmath}
\usepackage{graphicx}% Include figure files
\usepackage{dcolumn}% Align table columns on decimal point
\usepackage{bm}% bold math
\usepackage{epsfig}
\usepackage{hyperref}
\usepackage{soul,color}
\sethlcolor{black}
\def\figdir{}

\newcommand\figwidth{.48\textwidth}
\newcommand\Eq[1]{Eq.~\ref{eq:#1}}
\newcommand\Fig[1]{Fig.~\ref{fig:#1}}

\newcommand\Tab[1]{Table~\ref{tab:#1}}

\newcommand\calI{\mathcal I}

\newcommand\calL{\mathcal L}

\begin{document}

\preprint{%INT-XXX,
MIT-CTP/4691, JLAB-THY-15-2112, NT@WM-15-12}%, RBRC-XXXX, XXX-XXX}

\title{Massive Photons: An Infrared Regularization Scheme for Lattice QCD+QED}

\author{{Michael G. Endres}}
\email[]{endres@mit.edu}
\affiliation{{Center for Theoretical Physics, Massachusetts Institute of Technology, Cambridge, Massachusetts 02139, USA}}

\author{{Andrea Shindler}}
\email[]{a.shindler@fz-juelich.de}
\affiliation{{IAS, IKP and JCHP, Forschungszentrum J\"{u}lich, 52428 J\"{u}lich, Germany}}

\author{{Brian C. Tiburzi}}
\email[]{btiburzi@ccny.cuny.edu}
\affiliation{{Department of Physics, The City College of New York, New York, New York 10031, USA}}
\affiliation{{Graduate School and University Center, The City University of New York, New York, New York 10016, USA}}
\affiliation{{RIKEN BNL Research Center, Brookhaven National Laboratory, Upton, New York 11973, USA}}

\author{{Andr\'{e} Walker-Loud}}
\email[]{awalker-loud@lbl.gov}
\affiliation{{Department of Physics, College of William and Mary, Williamsburg, Virginia 23187-8795, USA}}
\affiliation{{Jefferson Laboratory, 12000 Jefferson Avenue, Newport News, Virginia 23606, USA}}
\affiliation{{Lawrence Berkeley National Laboratory, Berkeley, California 94720, USA}}

%\author{\hl{C.G.B. Spender}}
%\email[]{spender@area51.gov}
%\affiliation{\hl{Area 51, 134 km north-by-northwest of Las Vegas, NV 00051, USA}}

%\collaboration{Collaboration 51}%: X-Files: Project 0}


%\date{\today}

\begin{abstract}

Standard methods for including electromagnetic interactions in lattice quantum chromodynamics calculations result in power-law finite-volume corrections to physical quantities.
Removing these by extrapolation requires costly computations at multiple volumes.
We introduce a photon mass to alternatively regulate the infrared, and rely on effective field theory to remove its unphysical effects.
Electromagnetic modifications to the hadron spectrum are reliably estimated with a precision and cost comparable to conventional approaches that utilize multiple larger volumes.
A significant overall cost advantage emerges when accounting for ensemble generation.
The proposed method may benefit lattice calculations involving multiple charged hadrons, as well as quantum many-body computations with long-range Coulomb interactions.

\end{abstract}

\maketitle

{\bf Introduction --}
%\label{sec:intro}
%
Approximately $95\%$ of the visible mass of the Universe arises from the binding of quarks into nucleons by the strong interactions of quantum chromodynamics (QCD).
The relative mass difference between the proton and neutron is approximately $0.07\%$, and is attributed to two sources of isospin symmetry breaking in the standard model, namely, differences in the down and up quark masses and their electric charges.
Although these breaking effects are minute, they play an essential role in our understanding of the Universe.
For example, the primordial abundance of light nuclear elements in the early Universe is exquisitely sensitive to the excess mass of the neutron compared to the proton ~\cite{Weinberg:1102255,Walker-Loud:2014iea}.

Lattice QCD (LQCD) provides a first-principles approach for determining isospin-breaking effects in hadronic and nuclear processes.
There are a handful of LQCD calculations of the strong contribution to the nucleon mass splitting~\cite{Beane:2006fk,Blum:2010ym,Horsley:2012fw,deDivitiis:2013xla,Borsanyi:2013lga,Borsanyi:2014jba,Walker-Loud:2014iea}
and a comparable number that determine the electromagnetic corrections~\cite{Blum:2007cy,Basak:2008na,Blum:2010ym,Portelli:2010yn,Portelli:2012pn,deDivitiis:2013xla,Borsanyi:2013lga,Drury:2013sfa,Aoki:2012st,Ishikawa:2012ix,Horsley:2013qka,Borsanyi:2014jba}.
One impressive calculation includes both sources of isospin breaking simultaneously and yields, among other quantities, a postdiction for the nucleon isospin splitting with $\sim5\sigma$ statistical significance~\cite{Borsanyi:2014jba}.
There exists an alternate means for determining the electromagnetic self-energy of the nucleon, from the Cottingham formula~\cite{Cini:1959,Cottingham:1963zz,Gasser:1974wd,Collins:1978hi}, which makes use of experimental cross sections as input to dispersion integrals.
However, the uncertainty attained with this method~\cite{Gasser:1982ap,WalkerLoud:2012bg,Gasser:2015dwa} is not yet competitive with the LQCD calculations.

Although inclusion of electromagnetism in LQCD is theoretically straightforward~\cite{Duncan:1996xy,Duncan:1996sq}, it presents practical challenges due to the long-range nature of the electromagnetic (QED) interactions.
Specifically, such interactions give rise to power-law finite-volume (FV) corrections, and their removal via extrapolation requires computationally demanding simulations performed at multiple volumes.
An analytic understanding of the power-law FV effects within such setups~\cite{Davoudi:2014qua,Borsanyi:2014jba,Fodor:2015pna,Lee:2015rua} 
has enabled reliable FV extrapolations of the single hadron spectrum.

Despite the successful application of present techniques, there are a number of reasons for considering new methods.
Control over FV modifications to light nuclear binding energies seems to require particularly large volumes~\cite{Davoudi:2014qua}.
There are quantities in addition to the spectrum for which a precise knowledge of the QED modifications is needed, for example, corrections to hadronic matrix elements~\cite{Carrasco:2015xwa} and charged particle scattering~\cite{Beane:2014qha}, both of which suffer from infrared (IR) challenges.
LQCD calculations are performed with multiple ultraviolet (UV) regulators, providing valuable cross-checks on the continuum extrapolation of many important quantities~\cite{Aoki:2013ldr}.
Multiple IR regulators can do the same for LQCD calculations that include QED, but to date, only a few other formulations have been considered~\cite{Lehner:2015bga,Patella:2015aa,Lucini:2015hfa}. 
Of those, only one is constructed with a local quantum field theory~\cite{Patella:2015aa,Lucini:2015hfa}.
Finally, computationally efficient means of accounting for IR effects are always desirable, not just for lattice QCD+QED, but anywhere long-range Coulomb interactions are present (see, e.g., Ref.~\cite{PhysRevB.51.4014}).

\begin{figure}  
\includegraphics[width=\figwidth]{\figdir 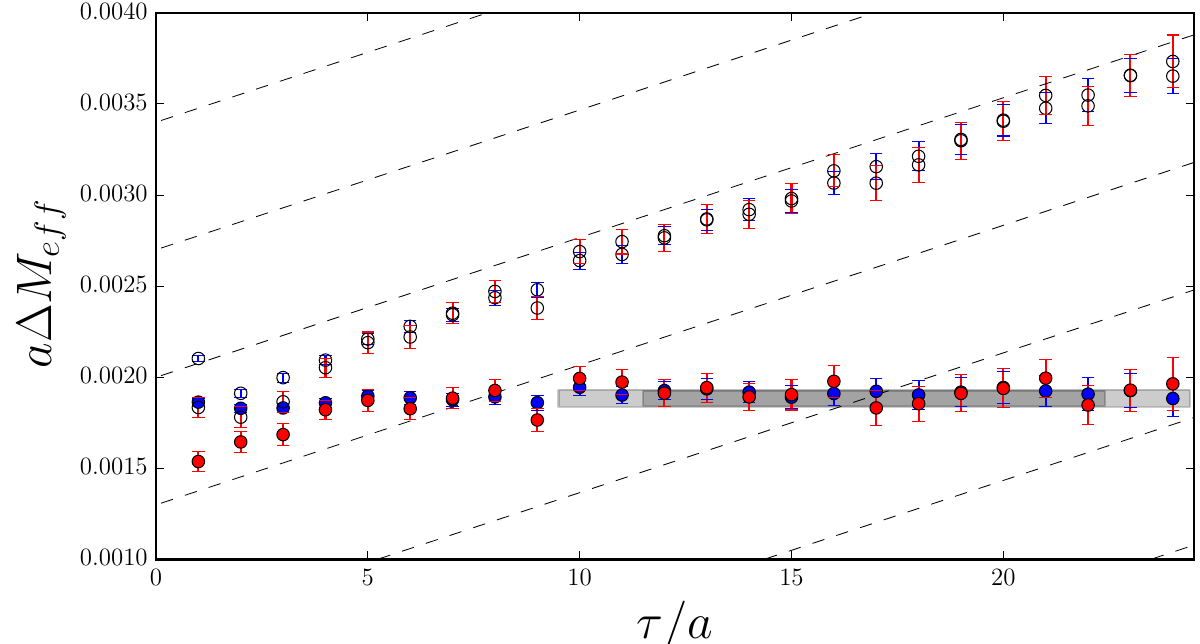}
\caption{\label{fig:meson_effective_mass_diff}%  
Zero-mode adjusted (filled) and unadjusted (open) effective mass difference for the kaon splitting ($m_\gamma/\bar m_\pi = 1/14$ and $L/a=24$).
The diagonal grid lines have slope $2x$; the red and blue points correspond to different sinks.
The gray bands correspond to uncertainties on the extracted value for $\Delta M_\textrm{eff}$.
} 
\end{figure}

Motivated by these considerations, we demonstrate the viability of an alternative IR regulator for lattice QCD+QED simulations: 
namely, the introduction of a photon mass $m_\gamma$.
Although a photon mass term manifestly violates gauge invariance, 
it maintains locality and
its effects on hadronic quantities can be reliably quantified and accounted 
for within an effective field theory (EFT) framework.
The introduction of a new scale, $m_\gamma$, implies an additional extrapolation within our approach.
With the aid of analytic formulas, however, we demonstrate that for the spectrum, such extrapolations can be performed {\it at a single volume} and yield results that are consistent with conventional approaches.
In the remaining sections, we present the salient features of our calculation.

{\bf Analytic considerations --}
%\label{sec:analytic}
%
In continuum Euclidean spacetime, the $R_\xi$ gauge fixed action for the massive photon is given by
\begin{eqnarray}
\calL_{\gamma} =  \frac{1}{4} F^2_{\mu\nu} + \frac{1}{2\xi} (\partial_\mu A_\mu)^2 + \frac{1}{2} m_\gamma^2 A^2_\mu 
\label{eq:noncompact_u1}
\end{eqnarray}
where $F_{\mu\nu} = \partial_\mu A_\nu - \partial_\nu A_\mu$; throughout this study, we work in Landau gauge, corresponding to the limit $\xi\rightarrow0$.
An Abelian theory, such as QED, with a massive vector gauge field is still perturbatively renormalizable.
This well known result follows from the fact that it is possible to find
a Becchi-Rouet-Stora-Tyutin (BRST) transformation that leaves the Lagrangian invariant up to a total divergence~\cite{Collins:1984xc}.
The BRST symmetry is preserved if one uses a gauge invariant UV cutoff~\cite{Luscher:1988sd}, such as a spacetime lattice; thus, the renormalizability follows from the power-counting theorems for a lattice regularization~\cite{Reisz:1987da}.

\begin{figure}  
\includegraphics[width=\figwidth]{\figdir 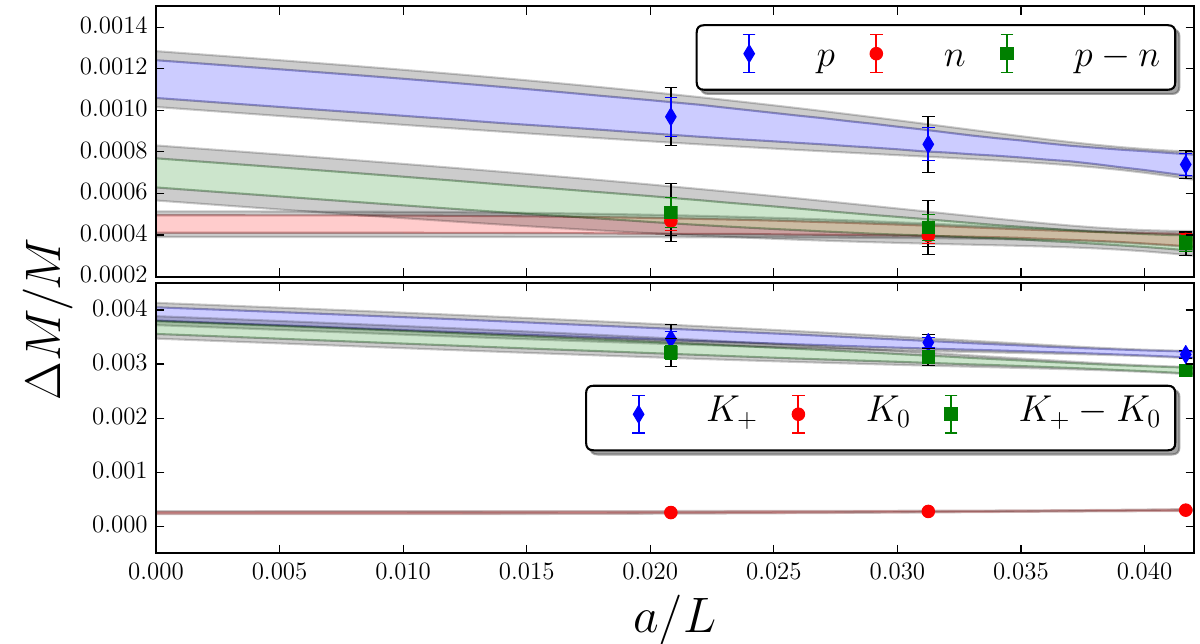}
\caption{\label{fig:nucleon_mass_diff_coulomb}%  
$\textrm{QED}_{TL}$ induced mass differences, extrapolated to infinite volume (taking $K_L=2$).
} 
\end{figure}

We consider three forms of corrections to correlators and hadron mass differences at leading order in the fine-structure constant $\alpha = e^2/(4\pi)$.
These corrections arise from either the zero mode contribution to the partition function, the presence of a finite photon mass, or FV effects.
The analytic forms of these corrections are determined from an EFT for hadrons of mass $M$ ($M=m_n$, $m_p$, $m_{K^+}$, and $m_{K^0}$) and charge $Q$; the naive expansion is in $m_\gamma / M$ (i.e. $\Lambda_\textrm{UV} = M$)~\footnote{Dynamical pions must be accounted for when $m_\gamma \gtrsim 2m_\pi$.
This can be handled within chiral perturbation theory but is not considered in the present work.}.
The EFT is a generalization of nonrelativistic QED~\cite{Caswell:1985ui} for hadrons that includes a photon mass term, and additional operators that are not constrained by gauge invariance. 

{\noindent (1) \it Zero mode:}
For sufficiently small 
$m_\gamma$, 
the zero mode of the temporal photon field appearing in \Eq{noncompact_u1} must be treated nonperturbatively \cite{Endres:2015vpi}.
In this regime,  
the two-point function for single hadrons has the form
\begin{eqnarray}
C(\tau) 
&=& Z e^{-M \tau-x\tau^2} \ ,
\label{eq:zero_mode}
\end{eqnarray}
where $Z$ is an overlap factor; the zero-mode contribution appears as $x=(4\pi\alpha Q^2)/(2m_\gamma^2 L^3 T)$, and vanishes as $T \to \infty$ at fixed $L$ and $m_\gamma$.

{\noindent (2) \it Photon mass:}
The hadron's electromagnetic mass shift can be determined as a function of photon mass, order by order in an expansion in powers of $m_\gamma / M$.
With the electromagnetic mass written as
$M(\alpha,m_\gamma)$, 
we define the mass shift
$\Delta_\gamma M(\alpha, m_\gamma) = M(\alpha, m_\gamma) - M(\alpha, 0)$,
which is UV finite. 
These IR shifts are given by
\begin{eqnarray}
\Delta_\gamma M^{LO} = -\frac{\alpha}{2} Q^2 m_\gamma, \cr
\Delta_\gamma M^{NLO} = \left(C e^2 - \frac{\alpha}{4\pi} Q^2 \right) \frac{ m_\gamma^2}{M} 
.\end{eqnarray}
The leading-order ($LO$) expression is nonanalytic in the squared photon mass, whereas the next-to-leading-order ($NLO$) expression is analytic but arises from both loops and local contributions~\footnote{The local operator contributing at $NLO$ is $\phi^*\phi \psi^\dagger \psi$, where the scalar $\phi$ picks up a vacuum expectation value, thereby Higgsing the photon.
Whereas the operator $m_\gamma^2 A_\mu A_\mu \psi^\dagger \psi$ contributes to masses at $N^3LO$, it is a prime example of a gauge non-invariant operator.};
the next-to-next-to-leading-order ($N^2LO$) correction is of order $\Delta_\gamma M^{N^2LO} = O(m_\gamma^3/M^2)$.
The latter two orders are accompanied by coefficients not fixed by the hadron charge.

{\noindent (3) \it Finite volume:}
The effects of FV can similarly be calculated using a nonrelativistive QED approach. 
This is a finite photon mass generalization of that pursued by Refs.~\cite{Davoudi:2014qua,Fodor:2015pna}.
The FV corrections to the electromagnetic mass are written as
$\delta_L M(\alpha,m_\gamma,L) = M(\alpha,m_\gamma,L) - M(\alpha,m_\gamma,\infty)$, 
and for charged hadrons are given up to $NLO$ by
\begin{eqnarray}\label{eq:mg_L_extrap}
\delta_L M^{LO}
&=& 
2\pi \alpha Q^2 m_\gamma 
\left[
\calI_1(m_\gamma L) 
- 
\frac{1}{(m_\gamma L)^3}
\right]
, \cr
\delta_L M^{NLO} 
&=& 
\pi \alpha Q^2 \frac{m_\gamma^2}{M} \left[ 2 \calI_{1/2}(m_\gamma L) + \calI_{3/2} (m_\gamma L) \right] , 
\end{eqnarray}
where
\begin{eqnarray}
\calI_n(z) = 
\frac{1}{2^{n+\frac{1}{2}} \pi^{\frac{3}{2}} \Gamma(n)} 
\sum_{\bm{\nu}\neq \bm{0}} \frac{K_{\frac{3}{2}-n}\left(z|\bm{\nu}|\right)}{ \left( z |\bm{\nu}| \right)^{\frac{3}{2}-n}} 
\end{eqnarray}
and $\bm{\nu} \in {\mathbb Z}^3$.
By contrast, the leading nonvanishing correction for neutral baryons [mesons] appears at $N^2LO$ [next-to-next-to-next-leading-order ($N^3LO$)].
Because the zero mode of the temporal photon is treated exactly in \Eq{zero_mode}, the FV corrections are calculated with this mode removed--a manifestation of which is the subtracted term appearing at LO.

\begin{table}
\caption{%
\label{tab:results_L_extrap}%
$\textrm{QED}_{TL}$ induced mass splittings, extrapolated to $L\rightarrow\infty$ ($m_\gamma =0$, $K_\gamma=0$).}
\begin{ruledtabular}
\begin{tabular}{c c c c}
splitting  &  $K_L$ & $\chi^2/\textrm{dof}$& $\Delta M/M \times10^{3}$ \\ 
\hline
$p-n$ & 1 & 0.07/2 & 0.73(05) \\
      & 2 & 0.03/1 & 0.70(13) \\
\hline
$K^+-K^0$ & 1 & 0.29/2 & 3.71(06) \\
          & 2 & 0.17/1 & 3.68(20) \\
\end{tabular}
\end{ruledtabular}
\end{table}

{\bf Lattice parameters and ensembles --}
%\label{sec:lattice}
%
Electroquenched numerical calculations of the hadron spectrum were performed using a modified version of the Chroma software suite \cite{Edwards:2004sx}.
Studies were performed using dynamical SU(3) flavor symmetric isotropic QCD gauge field configurations generated using a tadpole-improved L\"{u}scher-Weisz gauge action and clover fermion action.
The configurations correspond to a single lattice spacing $a = 0.1453(16)$~fm, three spatial extents: $L\sim3.48$ fm, $4.64$ fm and $6.96$ fm, and temporal extents $T>L$.
The pion (kaon) and nucleon masses in physical units are $\bar m_\pi = \bar m_{K} = 807.0(9.1)$~MeV and $\bar m_n = 1.634(18)$~GeV, respectively.
%The pion (kaon) and nucleon masses in physical units are $\bar m_\pi = \bar m_{K} = 807.0(0.2)(8.9)$~MeV and $\bar m_n = 1.634(0)(18)$~GeV, with uncertainties from a combined statistical and fitting systematic and lattice spacing, respectively.
%In lattice units these these masses correspond to a $a \bar m_\pi=\bar m_K = 0.5943(1)(1)$ and $a \bar m_n=1.2029(4)(2)$, respectively. 
%
This choice of masses ensures that the only appreciable FV corrections to hadron masses are those arising from QED effects.
The QCD ensembles used in this work comprise 956 ($L/a=24, T/a=48$), 515 ($L/a=32, T/a=48$) and 342 ($L/a=48, T/a=64$) configurations and are a subset of those described in Ref.~\cite{Beane:2012vq}; further details regarding the ensembles, lattice action and parameters can be found there.

\begin{table}
\squeezetable 
\caption{%
\label{tab:results_mg_extrap}%
$\textrm{QED}_{M}$ induced mass splittings, extrapolated to $m_\gamma\to 0$ ($L/a=24$, $K_L=1$).}
\begin{ruledtabular}
\begin{tabular}{c c c c c}
splitting &  $m_\gamma/m_\pi$ range & $K_\gamma$& $\chi^2/\textrm{dof}$& $\Delta M/M \times10^{3}$\\ 
\hline
$p-n$   & 1/14 - 1& 2& 0.09/5& 0.79(06)\\
	& 1/4 - 1/2& 1& 0.06/2& 0.81(08)\\
\hline
$K^+-K^0$& 1/14 - 1   & 2 & 0.42/5& 3.77(06)\\
         & 1/4 - 1/2  & 1 & 0.12/2& 3.79(06)\\
\end{tabular}
\end{ruledtabular}
\end{table}

The uncorrelated photon field configurations $A_\mu$ were generated using two different lattice actions: a conventional massless Coulomb gauge-fixed action with the zero mode removed ~\cite{Duncan:1996xy,Duncan:1996sq,Portelli:2010yn} ($\textrm{QED}_{TL}$)~\footnote{Note that this formulation of QED violates reflection positivity, leading to the ill-behavior of charged particle propagators for $T \to \infty$ at fixed $L$ \cite{Borsanyi:2014jba}. Our $\textrm{QED}_{TL}$ computation uses $T/L \sim 1$, for which the FV effect is mild.}, and a naive lattice discretized form of \Eq{noncompact_u1} ($\textrm{QED}_{M}$), where the derivatives are replaced by finite differences.
Note that in Euclidean space, Landau gauge is a complete gauge-fixing condition, and therefore in the latter case, the path integration over nonzero modes is well defined in the $m_\gamma\to0$ limit.
The photon mass values considered in this work are given by $m_\gamma/\bar m_\pi \in [1/14,1/7,1/4,1/3,5/12,1/2,7/12,1]$.
In both cases, results were obtained by computing correlation functions on QCD+QED gauge configurations generated by postmultiplying each QCD configuration by a single $e^{i e Q_q A_\mu}$, where $Q_u=2/3$, $Q_d=Q_s=-1/3$.
In the electroquenched approximation with SU(3) flavor symmetry, isospin splittings have missing contributions that are $O(\alpha^2)$, and therefore negligible for this study.

\begin{figure*}  
\includegraphics[width=0.45\textwidth]{\figdir 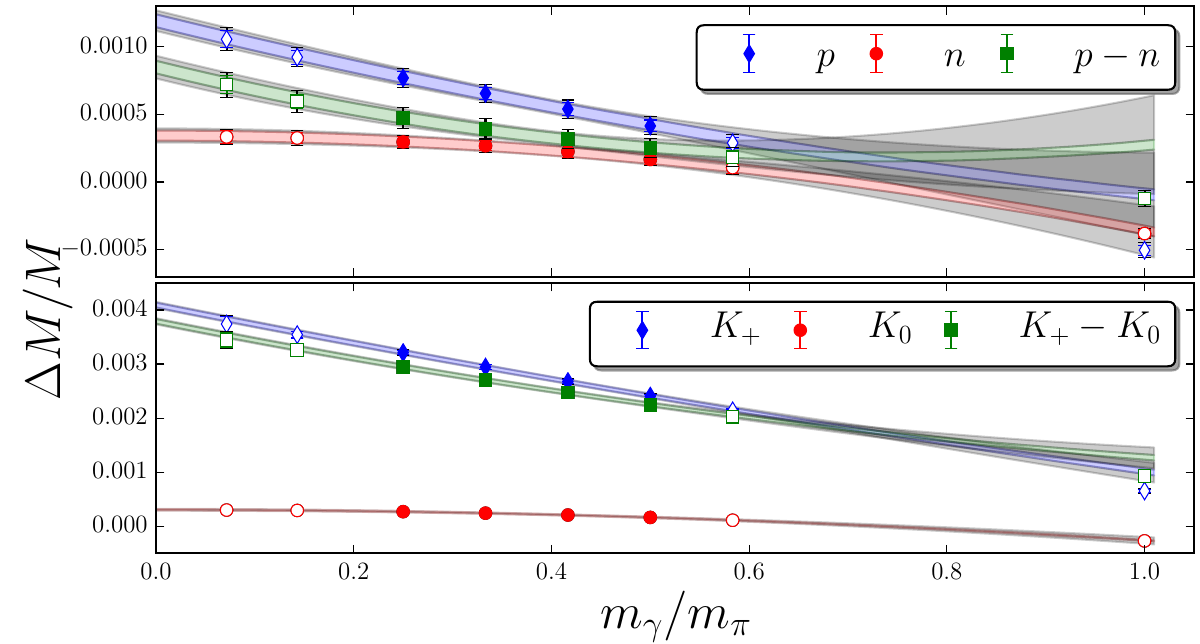}
\hspace{5pt} 
\includegraphics[width=0.45\textwidth]{\figdir 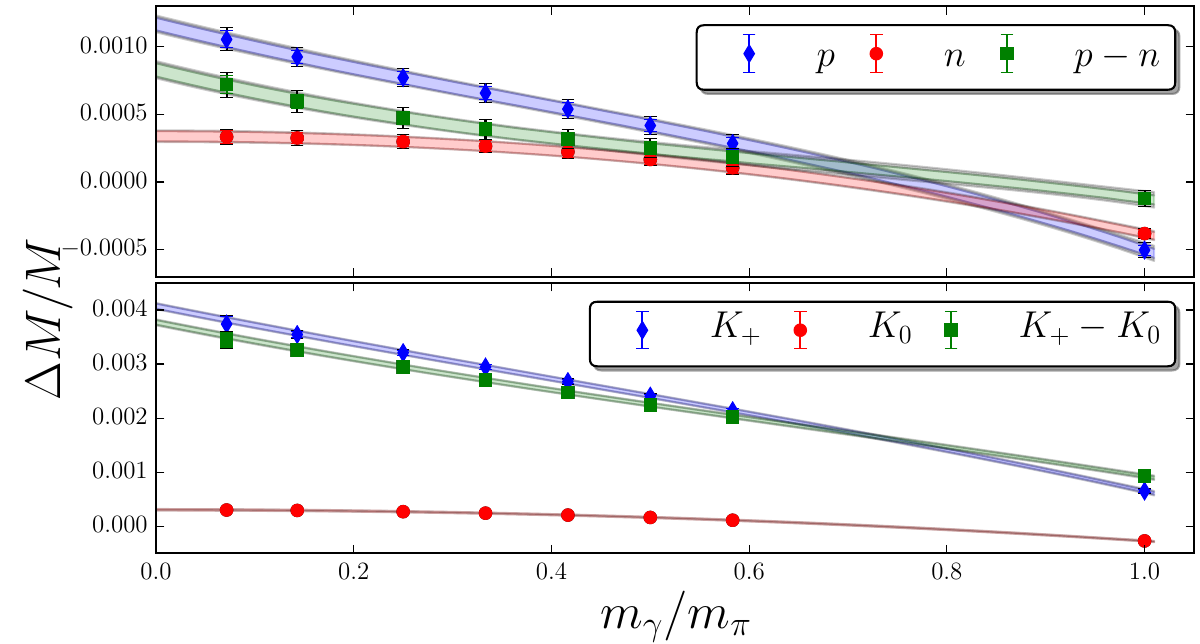} 
\caption{\label{fig:nucleon_mass_split_shift}%  
Volume adjusted ($K_L=1$) $QCD_M$ induced  mass differences, extrapolated to $m_\gamma\to0$. 
Fits were performed using $L/a=24$ data at all (right) and only the middle four (left) values of $m_\gamma$.
} 
\end{figure*}

In the electroquenched theory, the fine structure coupling does not renormalize
and therefore we take it to be equal to its experimental value $\alpha^{-1}=137.036\ldots$, 
measured in the Thomson limit.
The presence of electromagnetic interactions demands renormalization of the valence bare quark masses $m_q$, however.
Since our lattice regulator breaks chiral symmetry, this leads to an additive shift in the quark mass.
We tune the valence quark masses so that, in the presence of electromagnetic interactions, the neutral $\bar q q$ meson mass $m_{qq}$ obtained from the connected part of the $\bar q q$ correlation function is sufficiently close to the pion (kaon) mass $\bar m_\pi$.
For our electroquenched calculation, this choice of renormalization is robust but the quark mass renormalization in the full QCD+QED does not allow for a unique separation of the QED and QCD effects~\cite{Gasser:2003hk}.
All measurements were performed using valence quark masses $a m_u = - 0.25501$ and $a m_d=a m_s = -0.24750$ (the QCD bare quark mass is $a m_q = -0.2450$); 
the resulting mistuning for the charge neutral mesons was $\Delta m_{qq}/\bar m_\pi \lesssim 0.1\%$ for all values of $m_\gamma/\bar m_\pi \le 1$, where $\Delta m_{qq} \equiv m_{qq} - \bar m_\pi$.

The mistuning from strong isospin breaking can be estimated using chiral symmetry.
For the kaon, one finds 
\begin{equation}
\frac{\Delta m_{K^+ - K^0}}{\bar{m}_K} \simeq 
	\frac{1}{2} \frac{\Delta m_{uu} - \Delta m_{dd}}{\bar{m}_K}
	\lesssim 0.0004\, ,
\end{equation}
while the nucleon correction is given by
\begin{equation}
\frac{\Delta m_{n -p}}{\bar m_n} \simeq 
	\alpha_{d-u} \frac{2(\Delta m_{dd} - \Delta m_{uu})}{\bar m_\pi} \frac{\bar m_\pi^2}{4\pi f_\pi \bar m_n}\, .
\end{equation}
We can estimate the parameter $\alpha_{d-u}$ from the LQCD determination of the $m_d - m_u$ contribution to the nucleon mass splitting~\cite{Beane:2006fk,Blum:2010ym,Horsley:2012fw,deDivitiis:2013xla,Borsanyi:2013lga,Borsanyi:2014jba,Walker-Loud:2014iea} and find $\Delta m_{p-n}/\bar m_n \lesssim 0.0002$.
In both cases, mistuning is a potentially sizable correction to our results, which affects both the $\textrm{QED}_{TL}$ and $\textrm{QED}_{M}$ determinations.
Although a precise quark mass tuning is required for practical applications, it is not needed in the present proof-of-principle study~\footnote{An additional mistuning effect arises because of the $T$-dependence of quark masses in $\textrm{QED}_{TL}$~\cite{Borsanyi:2014jba}. Addressing this is required for practical applications, or could be eliminated by using the $\textrm{QED}_{L}$ formulation.}.

{\bf Analysis and results --}
%\label{sec:results}
%
Shell-shell and shell-point correlation functions were estimated using a single measurement per configuration, with a randomly chosen spacetime source location.
Following Ref.~\cite{Blum:2007cy}, we average observables over $+e$ and $-e$ on each configuration in order to exactly cancel off the $O(e)$ contributions to statistical noise.
Mass differences due to electromagnetic effects can be determined from the late-time dependence of the single hadron correlation functions $C^A(\tau)$ and $C^B(\tau)$, by studying the plateau region of an {\it effective mass difference} $\Delta M^{AB}_\textrm{eff}(\tau) =  M^A_\textrm{eff}(\tau) -  M^B_\textrm{eff}(\tau)$.
By exploiting the correlations between $A$ and $B$, we are able extract a clear signal for the mass difference.
For the nucleons, we consider a generalized effective mass formula of the form:
\begin{eqnarray}
M_\textrm{eff,exp}(\tau) = -\frac{1}{a} \log\frac{C(\tau+a)}{C(\tau)} + 2x \tau + x a\ ,
\end{eqnarray}
which neglects the backward propagation of states on a lattice of finite temporal extent $T$.
For mesons, we account for the backward propagating state by considering a generalized effective mass formula of the form:
\begin{eqnarray}
M_\textrm{eff,cosh}(\tau) =  \frac{1}{a} \cosh^{-1}
 \left[ \frac{e^{h(\tau,a) } + e^{h(\tau,-a)} }{2} \right] - x T \, ,
\end{eqnarray}
where $h(\tau,a) = xa(a-T+2\tau) + \log [C(\tau+a)/C(\tau)]$.
Both formulas treat the zero mode of the temporal photon field appearing in \Eq{zero_mode} nonperturbatively (for neutral hadrons $x=0$ and these expressions reduce to their conventional forms).
Although this contribution is negligible compared to the hadron masses, for the lattice parameters considered it can be comparable in magnitude to the mass differences we wish to extract.
\Fig{meson_effective_mass_diff} provides an explicit example of the behavior of $\Delta M_\textrm{eff}(\tau)$ for the kaon mass splitting, computed both with and without the zero-mode contribution accounted for.

Mass differences were determined for all volumes and photon masses via a correlated constant least-squares fit to $\Delta M_\textrm{eff}$ in the plateau region, as demonstrated in \Fig{meson_effective_mass_diff}.
An analogous determination from exponential fits to a ratio of correlation functions yielded consistent results.
Systematic uncertainties were estimated by varying the region over which fits were performed, and all uncertainties were added in quadrature.
The extracted mass shifts were subsequently extrapolated to vanishing photon mass and/or the infinite-volume limit using the fit formula:
\begin{align}
\Delta M(\alpha, L, m_\gamma) =&\ \Delta M(\alpha) + \sum_{k=0}^{K_\gamma} \Delta_\gamma M^{N^k LO}(\alpha,m_\gamma) 
\nonumber\\
&+ \sum_{k=0}^{K_L} \delta_L M^{N^k LO}(\alpha,m_\gamma,L) ,
\label{eq:extrap}
\end{align}
where $K_\gamma$ and $K_L$ indicate the order of each extrapolation.
In the case of mass splittings, an appropriate linear combination of mass shift formulas was used.
Note that for the $\textrm{QED}_{TL}$ extrapolations, $K_\gamma=0$; the appropriate FV formulas for $\delta_L M^{N^kLO}$ retain $T$ dependence, and may be found in Ref.~\cite{Borsanyi:2014jba}.

\begin{figure}  
\includegraphics[width=\figwidth]{\figdir 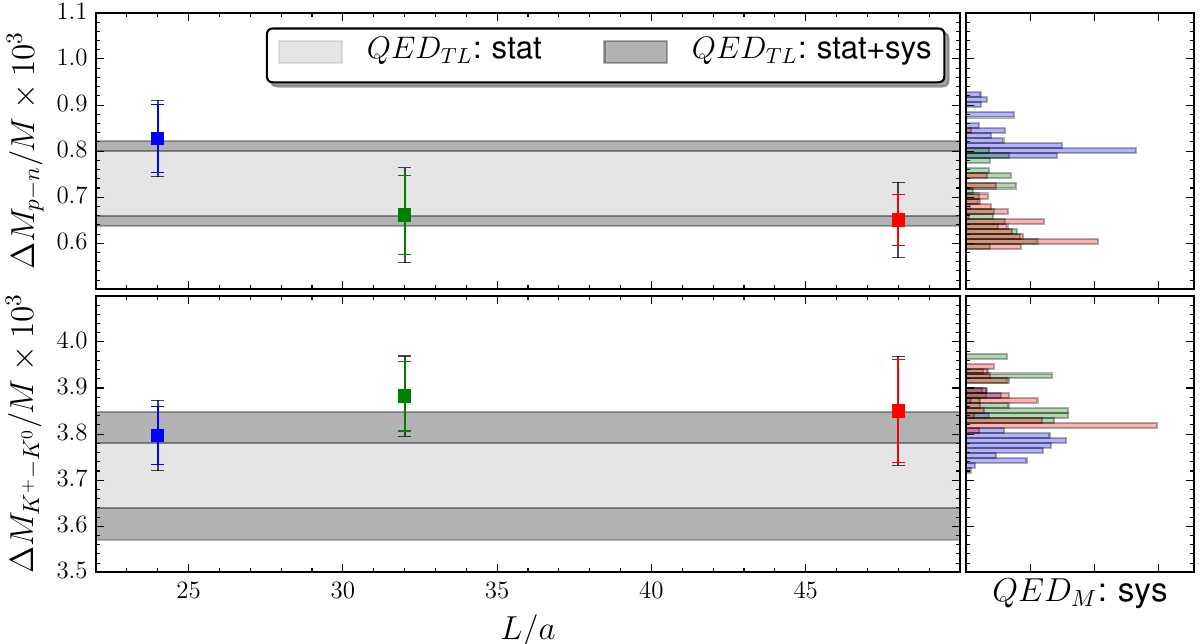}
\caption{\label{fig:histograms}%  
$\textrm{QED}_{M}$ extrapolations (points) performed independently at each volume.
Systematic errors are estimated by considering fits over multiple ranges of $m_\gamma$ and orders $K_\gamma\le3$ (summarized by the histograms).
$\textrm{QED}_{TL}$ extrapolations results are summarized by gray horizontal bands.
} 
\end{figure}

We carry out two independent analyses to test the viability of our proposal: (1) an infinite-volume extrapolation of $\textrm{QED}_{TL}$ induced mass differences, as is conventionally performed, and (2) an $m_\gamma\to0$ extrapolation of $\textrm{QED}_{M}$ induced mass differences using data at a {\it single} FV, but after having first removed the lowest order FV contributions $\delta_L M$.
Both types of extrapolation were performed using \Eq{extrap}, noting that many of the lowest-order contributions are fixed by theory.
Results for the first analysis, using all three volumes, are provided in \Tab{results_L_extrap} and representative fits are shown in \Fig{nucleon_mass_diff_coulomb}.
Results for the second analysis on the smallest volume are provided in \Tab{results_mg_extrap} for comparison, and shown in \Fig{nucleon_mass_split_shift}.
Analogous $\textrm{QED}_{M}$ extrapolations, performed at each of the three volumes, are summarized in \Fig{histograms} and are consistent not only with each other, but also the $\textrm{QED}_{TL}$ extrapolations.
In all cases, we find that the numerical and theoretical mass corrections are in excellent agreement down to at least $m_\gamma L \sim 1$.

The most computationally demanding part of our calculation involves multiple inversions of the Dirac operator.
Assuming, conservatively, a linear scaling with spacetime volume, the total inversion cost for $L/a=32$ is $515/956 \times(32/24)^3\sim1.3$ times greater than that of $L/a=24$.
By comparison, the $L/a=48$ inversion cost is $\sim3.8$ times greater.
The $L/a=24$ extrapolations using $m_\gamma/\bar{m}_\pi \in [1/4,1/2]$ data, provided in \Tab{results_mg_extrap}, are consistent with those using all values of $m_\gamma$.
The results are also consistent with the $\textrm{QED}_{TL}$ extrapolation using three volumes, provided in \Tab{results_L_extrap}, but required only $4/5$ the computational cost.
We therefore conclude that for the same precision and accuracy, the numerical cost of our $\textrm{QED}_{M}$ calculation of the mass splittings is comparable to or less than that of $\textrm{QED}_{TL}$.

{\bf Conclusion --}
%\label{sec:conclusion}
%
This work demonstrates that it is possible to reliably estimate infinite-volume hadron mass differences induced by electromagnetism on a single lattice volume with $\textrm{QED}_{M}$.
Conservatively, the pionless EFT employed in this work is valid for $m_\gamma \ll 2 m_\pi$ and $m_\pi L \gtrsim 4$.
Provided these inequalities are satisfied, the analytic expressions obtained for the mass shift are valid up to $O(m^3_\gamma/M^3,\alpha^2)$, and are independent of the pion mass; from our numerics, it appears that this order is sufficient to obtain reliable extrapolations of the mass shifts in the regime $m_\gamma / m_\pi \lesssim 1$ and $m_\gamma L \gtrsim 1$.

On preexisting lattice configurations, and for equal computational cost, we obtain an equally precise uncertainty in extrapolated differences as compared to the traditional method.  
This cost comparison does not account for the significant overhead of generating the configurations in the first place.
The results of our analysis pave the way for a more complete treatment of QED corrections using this approach.
When considering more involved LQCD calculations, such as charged-particle scattering~\cite{Beane:2014qha}, our method provides a mass gap to produce a photon, thus increasing the range of energy for which the standard L\"{u}scher method~\cite{Luscher:1986pf,Luscher:1990ux} for obtaining the scattering phase shift can be employed.
It will be interesting to explore these types of calculations, and also to use our method with chiral fermions, which do not suffer from additive quark mass renormalization.
Finally, it would be interesting to see if our method of screened interactions coupled with analytic extrapolation techniques is of benefit to quantum many-body calculations.

\begin{acknowledgments}
We would like to thank W.~Detmold, R.~Edwards, B.~Jo\'{o}, D.~Richards and K.~Orginos for the use of the JLab/W\&M QCD gauge field configurations and D.~B.~ Kaplan, T.~C.~Luu, and M.~J.~Savage for useful conversations and correspondences.
Additionally, we would like to thank A.~Patella and N.~Tantalo for stimulating discussions during the Lattice 2015 conference.
We acknowledge the hospitality of the International Institute of Physics at the Federal University of Rio Grande de Norte and the Institute for Nuclear Theory at the University of Washington (Nuclear Reactions Workshop~\cite{Briceno:2014tqa}), where portions of this work were completed.
Computations for this study were carried out on facilities of the USQCD Collaboration, which are funded by the Office of Science of the U.S. Department of Energy.
M. G. E. was supported by the U. S. Department of Energy Early Career Research Award DE-SC0010495, and moneys from the Dean of Science Office at MIT.
B. C. T. was supported in part by a joint City College of New York-RIKEN/Brookhaven Research Center fellowship, a grant from the Professional Staff Congress of the CUNY, and by the U.S. National Science Foundation, under Grant No. PHY15-15738.
A. W-L. was supported in part by U.S. Department of Energy (DOE) Contract No. DE-AC05-06OR23177, under which Jefferson Science Associates, LLC, manages and operates the Jefferson Lab and by U.S. DOE Early Career Award Contract No. DE-SC0012180.

\end{acknowledgments}

\bibliography{massive_qed}

\end{document}